\documentclass[epjCONF]{svjour}
\usepackage{graphicx}
\usepackage[varg]{txfonts} 
\usepackage[latin1]{inputenc}
\session-title{A Universe of Dwarf Galaxies}
\begin{document}
\title{Dwarf Galaxies in Clusters as Probes of Galaxy Formation and Dark Matter}
\author{Samantha J. Penny\inst{1}\fnmsep\thanks{\email{samantha.j.penny@gmail.com}} \and Christopher J. Conselice\inst{1}}
\institute{School of Physics \& Astronomy, University of Nottingham, Nottingham, NG7 2RD, United Kingdom}
\abstract{
We present the results of a \textit{Hubble Space Telescope (HST)} ACS and WFPC2 study of dwarf galaxies in the nearby Perseus Cluster, down to $M_{V} = -12$, spanning the core and outer regions of this cluster. We examine how properties such as the colour magnitude relation, structure and morphology are affected by environment for the lowest mass galaxies.  The low masses of dwarf galaxies allow us to determine their environmentally driven based galaxy evolution, the effects of which are harder to examine in massive galaxies. The structures of our dwarfs in both the core and outer regions of the cluster are quantified using the concentration, asymmetry and clumpiness (CAS) parameters.   We find that, on average, dwarfs in the outer regions of Perseus are more disturbed than those in the cluster core, with higher asymmetries and clumpier light distributions. We measure the $(V-I)_{0}$ colours of the dEs, and find that dwarfs in both the inner and outer regions of the cluster lie on the same colour magnitude relation. Based on these results, we infer that the disturbed dwarfs in the cluster outskirts are likely ``transition dwarfs'', with their colours transforming before their structures. Finally, we infer from the smoothness of the cluster core population that dwarfs in the inner regions of the cluster must be highly dark matter dominated to prevent their disruption by the cluster potential. We derive a new method to determine the minimum mass the dwarfs must have to prevent this disruption without the need for resolved spectroscopy, and determine their mass-to-light ratios. At their orbit pericentre, dwarfs in the core of Perseus require mass-to-light ratios between 1 and 120 to prevent their disruption, comparable to those found for the Local Group dSphs.} 
\maketitle
\section{Introduction} 
\label{intro}
Dwarf elliptical (dE) galaxies ($M_{\rm{B}} > -18$), and the fainter dwarf spheroidal (dSph) galaxies ($M_{\rm{B}} > -14$), are the most numerous galaxy type in the Universe. As such, any galaxy evolution/formation scenario must clearly be able to predict and describe the properties of these galaxies. As well as being numerous, dwarfs have low masses ($<10^{9}$ M$_{\odot}$), and are therefore more easily influenced by their environment than more massive galaxies, with environment likely playing a vital role in their formation and evolution. 

One well-known environmental dependence that dwarfs follow is the morphology-density relation defined by giant galaxies \cite{dressler80}, such that the fraction of early type galaxies increases with increasing environmental density. It was noted that dwarf irregulars were not found in great numbers in the Virgo Cluster \cite{reaves62} , despite such galaxies being numerous in the Local Group, providing the first hints that an environmental dependence of dwarf galaxy morphology exists. The morphology-density relation was subsequently shown to extend to the dwarf galaxy regime \cite{binggeli87}, with the number fraction of dwarf ellipticals increasing with local projected density. This relationship hints that the formation of dwarfs may be linked to the environment in which they reside. However, this environmental dependence of dwarf evolution is complicated by the fact that dwarfs in the cores of rich clusters are not a homogeneous population, showing a range of ages and metallicities  \cite{poggianti01,me08}.

It is hypothesized that the spread in ages and metallicities can be explained if at least some dwarfs are the result of the morphological transformation of infalling galaxies. Several formation scenarios have been proposed to explain the infall origin of such cluster dwarf ellipticals, with a number of environmental processes thought to influence the formation and evolution of these galaxies. Dwarf galaxies are likely susceptible to high speed encounters with other galaxies (harassment \cite{moore}), ram pressure stripping by the hot ISM \cite{grebel03}, or by interactions with the tidal potential of their host cluster, group, or galaxy. These processes can cease star formation and remove mass from an infalling progenitor galaxy, transforming it from late to early type. The efficiency of these processes depend on the density of the environment in which the dwarf resides. However, the degree to which environment  shapes these low mass galaxies is largely unknown.

In this paper we compare the properties of dwarf galaxies in both the core and outskirts of the  Perseus Cluster. Using deep \textit{Hubble Space Telescope (HST)} Advanced Camera for Surveys (ACS) and Wide Field Planetary Camera 2 (WFPC2) observations in the F555W and F814W bands ($V$ and $I$ respectively), we target a total of twelve fields in the cluster, spanning a range of cluster-centric distances, to examine the effect of local environmental density on the cluster dwarf galaxy population. We examine the structures of the dEs in the cluster outskirts and compare them to those in the cluster core, to look for any evidence of an infall origin for the dwarfs. Dwarfs in the cluster outskirts are, on average, more disturbed than those in the cluster core, which we interpret as evidence for a recent infall origin for these dwarfs. Despite being more disturbed than those in the cluster core, these dwarfs show no evidence for recent star formation, following the same colour magnitude relation as those in the cluster core. This result suggests that these dwarfs have recently transformed from late-type to early-type galaxies, with their colours transforming prior to their structures. 

\section{Data and Observations}
\label{sec:obs}
We examine the dwarf galaxy population of the Perseus Cluster (Abell 426), one of the nearest rich galaxy clusters, with a redshift $v = 5366$ kms$^{-1}$ \cite{Stublerood99}, and a distance $D = 72$ Mpc. To examine how local environmental density shapes the cluster dwarf galaxy population involves deep, high resolution imaging sampling the whole cluster, from its dense core to sparse outer regions. We therefore utilise $HST$ ACS and WFPC2 imaging in the F555W and F814W bands. We target five fields in the central 300 kpc $\times$ 300 kpc of the cluster core using ACS, and seven fields in the cluster outskirts using WFPC2, out to a distance of 600 kpc from the cluster centre. WFPC2 was used due to the failure of ACS. Our survey design is shown in Fig. 1. 

\begin{figure}
\includegraphics[width=75mm]{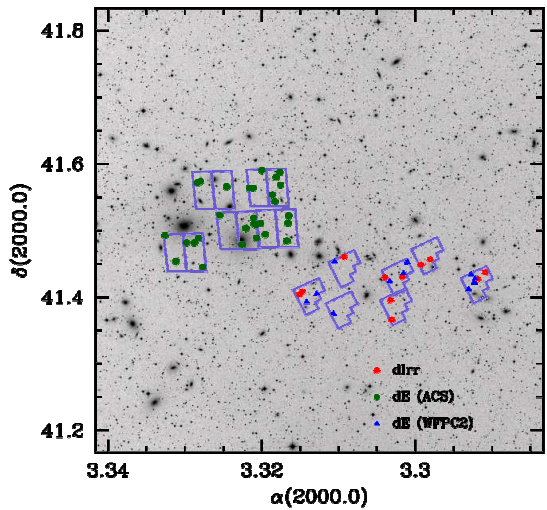}
\caption{Positions on the sky of the five $HST$/ACS (double rectangles) and seven $HST$/WFPC2 (chevrons) pointings, covering both the core and outer regions of the Perseus Cluster. Also marked are the postions of all dwarf ellipticals identified in this study. The fields and dE positions are overplotted on a DSS image spanning 600 kpc $\times$ 600 kpc of the cluster. }
\label{survey}
\end{figure} 

\subsection{Cluster Membership}
\label{sec:clustermem}
The high resolution of $HST$ imaging allows for the separation of cluster galaxies based on sizes, structures and morphologies \cite{me09,trent09}. Dwarf ellipticals are diffuse and symmetric in appearance, whereas background galaxies often contain spiral structures or tidal features. Dwarf ellipticals also have low surface brightness, whereas background galaxies do not. Furthermore, dEs can be separated from background ellipticals as the light distributions of dEs are less concentrated. This method was utilised for our ACS imaging of the cluster core \cite{me09}, resulting in a sample of 29 dEs in the cluster core. We use this method for identifying cluster dwarfs in our WFPC2 imaging, resulting in a sample of 11 dEs in the cluster outskirts.

\section{Structures and Morphologies}
\label{sec:structures}
It is hypothesized that some dwarf ellipticals are formed via tidal processes such as ram pressure stripping and the harassment of galaxies infalling into the cluster \cite{moore}. These transformations are thought to take place in environments with low velocity dispersions, such as galaxy groups which are later accreted by the cluster. The outskirts of a rich cluster represent a region where galaxies that have been recently accreted into the cluster are likely to reside. If the transformation from a late type galaxy to an early type dwarf elliptical does indeed take place in groups that are later accreted by the cluster, we might expect to find in the cluster outskirts a population of dwarfs that retain evidence of this transformation, such as remnant bars or tidal features \cite{sven02}.

Previous studies have found example objects, termed transition dwarfs \cite{grebel03}, which are galaxies that show characteristics of both early-type dwarfs (dEs \& dSphs) and late-types (dIrrs). It is likely that dIrrs are the precursors of dEs, with dEs being dIrrs that have lost their gas or converted it into stars long ago \cite{kormendy85,kormendy09}. Transition dwarfs likely represent a class of dwarfs undergoing the evolutionary transformation between late and early type galaxies, and may provide important clues as to the origin of dwarf ellipticals/spheroidals. These galaxies will have morphologies intermediate between dIrrs/dEs \cite{sandage91}, and we therefore examine the structures of the dwarfs identified in our WFPC2 imaging to identify such objects.

\subsection{Concentration, Asymmetry and Clumpiness Parameters}  
\label{sec:cas}
We quantify the structures of our newly identified dwarfs in the outer regions of Perseus using the concentration, asymmetry and clumpiness (CAS \cite{cas}) parameters. These parameters are measured using the inverted form of the Petrosian radius \cite{Pet76,Kron95,me09} for each galaxy. Each dwarf is manually cleaned of background galaxies and foreground stars using the {\sc{iraf}} task IMEDIT prior to the calculation of the CAS parameters, as these could masquerade as features in the dwarfs, or contaminate the measurements.

We compute the concentration index (C) for each galaxy, which is defined as the ratio of the radii containing 80$\%$ and 20$\%$ of a galaxy's light. The higher the value of C, the more concentrated the light of the galaxy is towards the centre, with dwarf ellipticals having an average value of 2.5 $\pm$ 0.3, compared to 4.4 $\pm$ 0.3 for giant ellipticals \cite{cas}. The asymmetry indices (A) for the dwarfs in our sample are also computed. This index measures the deviation of the galaxies light from perfect 180$^{\circ}$ symmetry. The less symmetric the galaxy, the higher the asymmetry index. Asymmetric light distributions are produced by features such as star formation, galaxy interactions/mergers and dust lanes \cite{Conselice00}, which are not found in dwarf ellipticals in the core of Perseus \cite{me09}. Early-type galaxies such as dwarf ellipticals are expected to have average asymmetries that are near zero, compared to $<$A$>$ $= 0.17$ for irregulars. 

The third index we measure is the clumpiness index (S). This index describes how patchy the distribution of light within a galaxy is. The clumpiness (S) index is the ratio of the amount of light contained in high frequency structures to the total amount of light in the galaxy. For dwarf ellipticals in the core of Perseus, this is found to be near zero. However, if dwarf ellipticals in the outskirts of Perseus have only recently been transformed from late to early-type galaxies, then we might expect some internal substructure to remain. Galaxies with internal sub-strcutre will have higher values of S, with $<$S$>$ $= 0.4$, typical for irregular galaxies \cite{cas}. This is because such galaxies typically contain young, blue stars have short life-times ($<100$ Myr \cite{bruzualcharlot03}) compared to the time-scale for star clusters and large scale star forming regions to dissolve ($> 1$ Gyr \cite{hasegawa08}).

We compare the CAS values for dwarfs in the core and outer regions in Fig.~\ref{casr}. Dwarf ellipticals in the outskirts of Perseus have higher values of both A and S than those in the cluster centre, with dwarfs in the outer regions of the cluster having asymmetry and clumpiness values $<$A$>$ $=0.13\pm0.09$, and $<$S$>$ $=0.18\pm0.08$, compared to $<$A$>$ $=0.02\pm0.04$, $<$S$>$ $=0.01\pm0.07$ for those in the cluster core. The errors quoted are 1$\sigma$ from the mean. 

The concentration index C does not reflect irregularities in a galaxy's light distribution, but we compare the results for the two samples for completeness. We find that C is consistent between the two environments, with $<$C$>$ $= 2.58 \pm 0.29$ for dwarfs in the cluster outskirts, compared to $<$C$>$ $= 2.52 \pm 0.21$ for those in the cluster core. Therefore, within the error bars, the two populations have the same concentration values.

 \subsection{Environmental Dependence of Morphology}
\label{sec:endep}
Local density decreases with increasing cluster-centric radius, such that the cores of clusters are more densely populated than the cluster outskirts. Galaxy morphology is dependent on local galaxy density, with spiral and irregular galaxies more commonly found in regions of low local density (the morphology-density relation, \cite{dressler80,binggeli87}). Our $HST$ survey areas are not large enough to accurately determine local galaxy density, so we instead use projected cluster-centric distance as an indicator for local galaxy density. 

\begin{figure*}
\includegraphics[width=160mm]{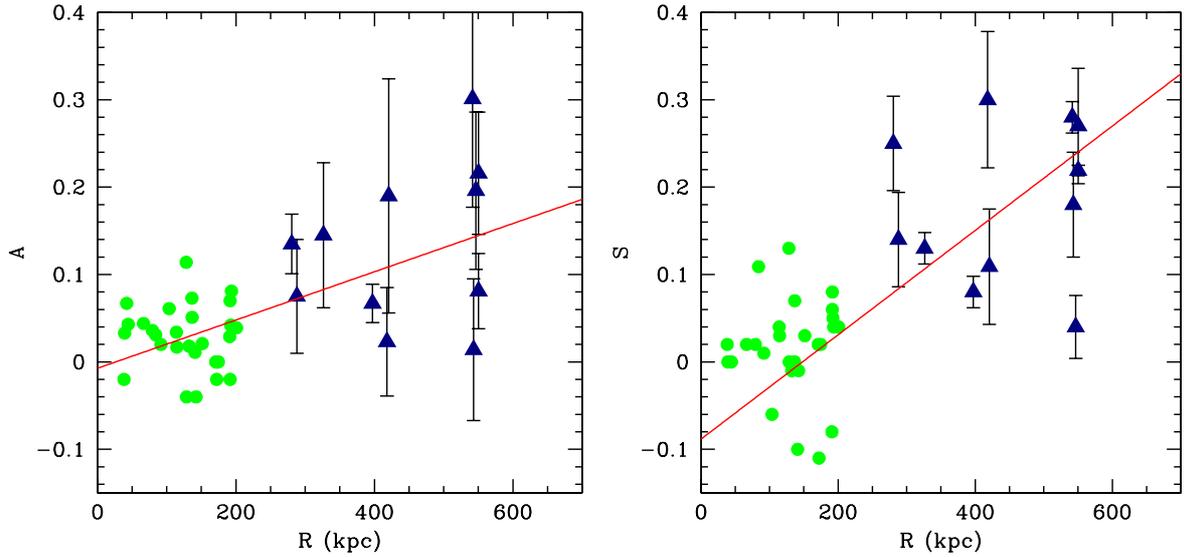}
\caption{Plots of A versus R (projected distance from the cluster center) and S versus R for the dwarfs in our sample. The circles are dwarfs in the core of the cluster, and the triangles are dwarfs in the outer regions. The red lines are least-squares fits to the data.}
\label{casr}
\end{figure*}

To determine how the morphologies of our dwarfs change with cluster-centric distance, we compute the projected distance of each dwarf from the cluster centre (taken to be the cluster centred galaxy NGC 1275), and compare these distances to their A and S values. The values of C are not investigated as a function of environment, as this index provides a useful way of separating cluster dEs from background ellipticals, but does not give a measure of features due to recent star formation or interactions \cite{cas}. The results are shown in Fig.~\ref{casr}. 

We fit the relationship between both A and S with R using a weighted least squares fit, using bootstrap samples. We generate 10,000 bootstrap samples, each of the same size as the original data set. A weighted least squares fit was then performed to each of the bootstrap samples. We then find the mean and dispersions of these fits, with these best fit lines shown as red lines on Fig.~\ref{casr}. The errors quoted below are the dispersions of the least squares fits to the bootstrap samples. The following fits to the data are found:
\begin{equation}
A = (0.028\pm0.010)\left(\frac{R(kpc)}{100 kpc}\right) - (0.007\pm0.018),
\end{equation}
\begin{equation}
S = (0.060\pm0.024)\left(\frac{R(kpc)}{100 kpc}\right) - (0.088\pm0.053),
\end{equation}
where $R$ is the projected distance from the cluster centre. There is a correlation between both $A$ and $S$ with projected cluster centric distance $R$, such that the more distant a dwarf is from the cluster centre, the more disturbed its morphology is.

\section{Colour Magnitude Relation}
\label{sec:CMR}

A well-defined colour-magnitude relation (CMR) exists for elliptical galaxies, such that the brightest galaxies are the reddest in colour, as first noted for ellipticals in the Virgo and Coma clusters \cite{sandage72}. This correlation is generally regarded as a relationship between a galaxy's mass (as traced by luminosity), and metallicity (traced by colour \cite{smithcastelli,sven09}). Dwarf galaxies follow this same colour magnitude relation \cite{me08,sven09}. Using the newly identified dwarfs in the outer regions of Perseus, we investigate whether dwarfs in the lower density regions of rich clusters lie on this same colour-magnitude relation. Given that the morphologies of the dwarfs in the outskirts of Perseus suggest they have been recently transformed from late to early type, their colours may also reflect this, in that a recently transformed galaxy might be expected to be bluer in colour due to having a younger stellar population.

\begin{figure}
\includegraphics[width=75mm]{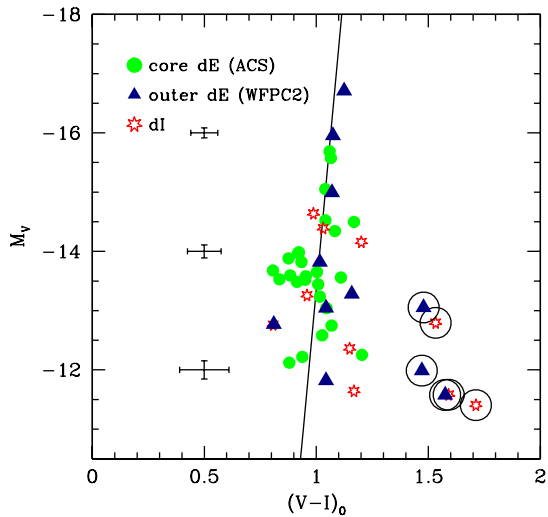}
\caption{Colour magnitude relation for dwarf ellipticals in Perseus. The filled circles represent dwarfs in the core of Perseus (ACS), the triangles represent those in the outer regions of the cluster (WFPC2), and the stars are dIrrs. The solid line is a linear fit to the colour magnitude relation for the dwarf elliptical population of Perseus as a whole, excluding any galaxies with ambiguous cluster membership (circled points, dwarfs 2, 3 and 7).}
\label{CMD}
\end{figure}

Our colour-magnitude relation is shown in Fig.~\ref{CMD}. We compare the CMR for the core and outer region dwarf populations, to examine any environmental dependence on the CMR. The CMRs for the two regions are fitted using a least squares method with the points weighted according to the error on their colour. This provides the following best fits to the data:

\noindent Core:
\begin{equation}
(V-I)_{0} = (0.76\pm0.30)-(0.017\pm0.022)M_{V},
\end{equation} 
\noindent and for the outskirts: 
\begin{equation}
(V-I)_{0} = (0.66\pm0.23)-(0.027\pm0.016)M_{V}.
\end{equation}Galaxies redder than $(V-I)_{0} = 1.3$ are excluded from this fit, due to the ambiguity of their cluster membership (circled points on the CMR)

Within the error bars, the colour magnitude relations for the two environments are identical. This suggests that given the disturbed morphologies of the dwarf galaxy population the outskirts of Perseus, it appears that the colours of dwarfs transform prior to their morphologies. If this were not the case, we would expect the dwarf population in the outer regions to be significantly bluer than those in the core, with their colours matching better their disturbed morphologies. The dwarfs in both the cluster core and outskirts have colours consistent with passive stellar populations that stopped forming stars long enough ago for there to be no age information left \cite{smithcastelli,sven09}.

\section{The dark matter content of the cluster core population}
\label{sec:core}

Down to our surface brightness limit, the dwarfs in the cluster core are surprisingly smooth in appearance, lacking in tidal features. Based on the smoothness of our sample of dEs, we can infer that these objects have a large dark matter component to prevent their tidal disruption by the cluster potential. However, the large distance to the Perseus Cluster prevents us from easily determining mass-to-light ratios for these dwarfs via spectroscopy. Therefore, we derive a new method for determining the dark matter content of these dwarfs by finding the minimum mass the dwarfs must have in order to prevent tidal disruption by the cluster potential.We calculate the minimum masses these dwarfs must have to prevent their disruption by the cluster tidal potential based on their sizes, projected distance from the cluster centre and the mass of the cluster interior to each dwarf. The minimum mass the dwarfs must have to prevent their disruption, $M_{dwarf}$ is calculated as:
\begin{equation}
M_{dwarf} > \frac{r_{d}^{3}M_{cl}(R)(3+e)}{R^3}
\end{equation}
\noindent where $r_{d}$ is the dwarf's Petrosian radius, $M_{cl}$ is the mass of the cluster interior to the dwarf, and $R$ is the projected distance of the dwarf from the cluster centred galaxy NGC1275.  The mass of the cluster interior to each dwarf is calculated using a model of the acceleration due to gravity in the core of Perseus provided by Mathews et al. 2006 \cite{Mathews06}, based on X-ray observations of the Perseus Cluster.

At its  pericentre, a dwarf will be subject to the largest tidal forces it will experience during its orbit, and therefore this will be the point at which its mass will need to be highest to prevent tidal disruption. We find that no dwarfs down to M$_{\rm{B}} = -12.5$ exist within a 35 kpc radius of NGC 1275. Therefore we take 35 kpc to be the pericentric, and thus minimum, distance from the cluster centre at which a dwarf can survive the cluster potential. 

\begin{figure*}
\includegraphics[width=160mm]{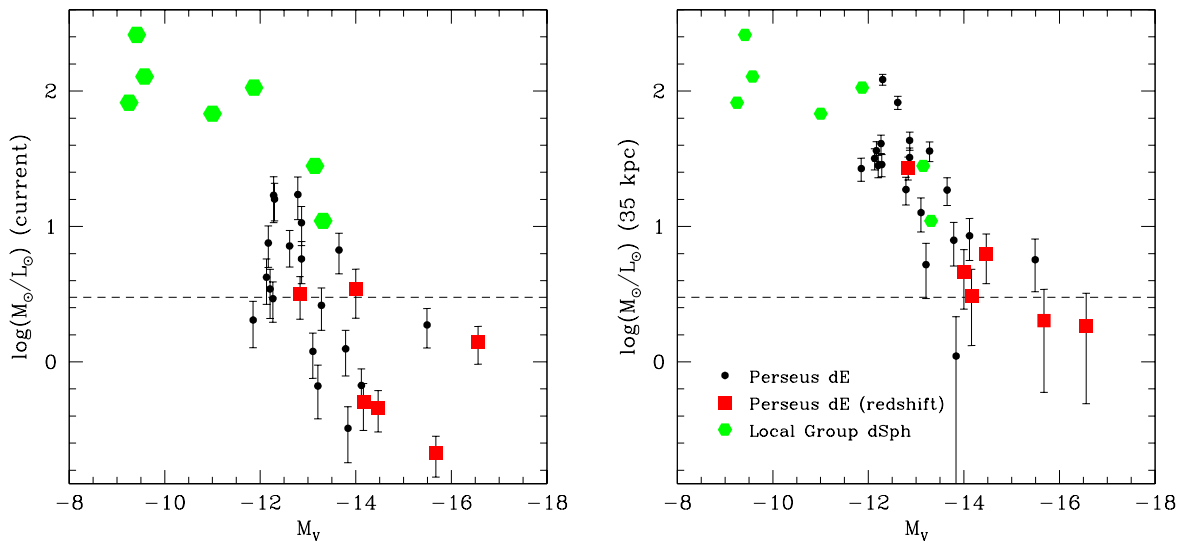}
\caption{Relationship between mass-to-light ratios and $M_{V}$ for Perseus dEs and Milky Way dSphs. The left hand plot shows M/L for the dwarfs at their current cluster position, and the right hand plot shows M/L for the dwarfs at a distance of 35 kpc from NGC 1275. Values of M/L for the Milky Way dSphs are taken from \cite{strigari07}, and were obtained using kinematics. The dashed line in each plot is at a M/L of 3, which we take, within our errors, to be the mass-to-light ratio at which no dark matter is required to prevent the dwarf being tidally disrupted.}
\label{m2l}
\end{figure*}

We calculate the minimum masses the dwarfs must have to prevent their disruption by the cluster tidal potential at both their current projected distances from the cluster centre, at at a pericentric distance of 35 kpc. These results are then compared to their luminous masses to obtain mass-to-light ratios for the dwarfs. These results are shown in Fig.~\ref{m2l}.

Figure~\ref{m2l} shows that at their current cluster positions, twelve of the dwarfs in our sample require dark matter, with the remaining dwarfs having mass-to-light ratios smaller than 3, indicating they do not require dark matter within our errors at their current distances from the cluster centre. When we measure the maximum  mass-to-light ratios the dwarfs could have, i.e. the mass they require to prevent disruption at a distance of 35 kpc from the cluster centre, we find that all but three require dark matter (Figure~\ref{m2l}b). There is a clear correlation between mass-to-light ratio and the luminosity of the dwarfs, such that the faintest dwarfs require the largest fractions of dark matter to remain bound. This is to expected, as the fainter a galaxy is, the less luminous mass it will contain, therefore the higher its dark matter content must be to prevent its disruption. 

\section{Discussion and Conclusions}
\label{sec:conc}

We identify from $HST$ ACS imaging a total of 11 dwarf ellipticals in the outskirts of the Perseus Cluster, all of which are newly discovered. We compare the colours, structures and morphologies of these newly discovered dwarfs in the cluster outskirts to those in the core region, for which we have $HST$ ACS imaging. To quantify their morphologies, we compute their concentration, asymmetry and clumpiness (CAS) parameters and compare these results to those obtained in \cite{me09} for the dwarfs in the core. The dwarfs in the cluster outskirts have, on average, more disturbed structures than those in the core, with $<$A$>$ $=0.13\pm0.09$ and $<$S$>$ $=0.18\pm0.08$, compared to $<$A$>$ $=0.02\pm0.04$, $<$S$>$ $=0.01\pm0.07$ for those in the cluster core.

All the dEs we identify in this study as definite cluster members lie on the colour-magnitude relation expected for dEs in all environments. If the more disturbed dwarfs originate from an infalling population of galaxies transforming from late-type to early-type, they have lost their gas and ceased star formation sufficiently long ago that their stellar populations resemble those of passive galaxies. We infer that these galaxies are ``transition dwarfs'' \cite{gallagherhunter} that lie on the red sequence but do not have the smooth structures found for the dwarf ellipticals in the cluster core. This can be explained by the timescales required to transform a galaxy's colour and structure. The process of removing gas from a star forming dwarf can take place over short timescales given the shallow potential wells of these galaxies. 

\begin{figure}
\includegraphics[width=82mm]{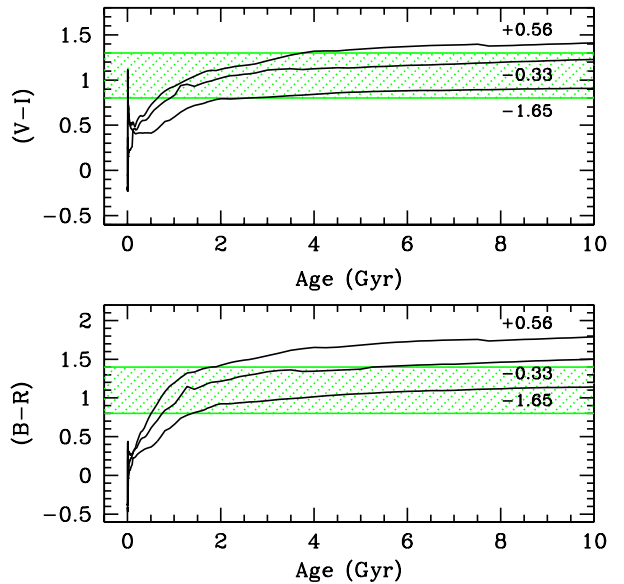}
\caption{The evolution of the \cite{bruzualcharlot03} models of the ($V-I$) and ($B-R$) colour indices with time. We include models for [Fe/H]$ = -1.65, -0.33,$ and $+0.56$. The shaded region represents the range in colours for our dwarfs in both environments, excluding the very red cluster candidates in our WFPC2 imaging due to the uncertainty in their cluster membership.}
\label{bruzualcharlot}
\end{figure}

The colours of star forming galaxies are dominated by young, blue O and B type stars. However, once the dwarf's star forming material has been removed by a process such as ram pressure stripping, the reservoir of gas required to fuel this star formation no longer exists. Therefore the colour of the galaxy will rapidly redden due to the short main-sequence life-times of massive stars. We can use the \cite{bruzualcharlot03} models to find over what timescale a galaxy will transform from blue to red once star formation ceases. We plot ($B-R$) and ($V-I$) colours versus time in Fig.~\ref{bruzualcharlot}. The ($B-R$) colours are included for reference as this colour will evolve faster than the ($V-I$) colour. It can be seen that a blue stellar population evolves rapidly to a red one, with a galaxy transforming from blue to red over a timescale of $\sim$0.5 Gyr.

To completely transform from a late to early type, internal sub-structure must be removed from the galaxy. We can estimate over what timescale star clusters within the dwarf will dissipate, removing substructure caused by star formation. Stars typically form in clusters of $M \sim 10^{3}$ M$_{\odot}$, and these will disperse and dissolve over time.  Such clusters in dense environments such as the center of the MW will dissolve rapidly, on a timescale of $\sim10$ Myr \cite{kim00}. However, dwarf galaxies typically have stellar densities much lower than in more massive galaxies, and therefore the timescale for star clusters to evaporate will be longer. Open clusters in the outer regions of the MW are found to have ages between 0.5-6 Gyr (e.g. \cite{hasegawa08}). It is likely therefore that structure will remain in a dwarf galaxy that has ceased star formation for at least a Gyr, though detailed simulations will be required to obtain a more precise life-time. Therefore, the morphological transformation from late to early type galaxy will take longer than the colour transformation.

Given their low masses, we might expect dwarfs in the core of the cluster to show evidence for disruption by the cluster tidal potential. To prevent this disruption, the dwarfs in the cluster core are likely highly dark matter dominated to explain their remarkably smooth structures. At their current projected distances from the cluster core,  12 out of 25 of the dwarfs in our sample require dark matter to remain stable against the cluster potential. At a pericentric distance of 35 kpc from the cluster centre, the mass-to-light ratios of these dwarfs are comparable to those of the Local Group dSphs, ranging between M$_{\odot}$/L$_{\odot}$ $\approx 1$ and 120.

\end{document}